Using Mobile Devices to Augment Inquiry-Based Learning Processes with Multiple

Representations

Sebastian Becker[a], Pascal Klein[a], Alexander Gößling[b], Jochen Kuhn[a]

[a] University of Kaiserslautern, Physics / Physics Education Research Group (Kaiserslautern,

Germany)

[b] Institute for School, Educational and Professional Science (Bielefeld, Germany)

**Corresponding author:**

Sebastian Becker

Erwin-Schrödinger-Str. 1

67663 Kaiserslautern

email-address: s.becker@physik.uni-kl.de

ORCID: 0000-0002-2461-0992

Using Mobile Devices to Augment Inquiry-Based Learning Processes with Multiple

Representations

## 1 Introduction

Multimedia learning research has a long tradition in educational psychology in general and in science education in particular. The main focus of this research has been desktop-based or virtual learning environments in science education, especially with animations and simulations (e.g. Makransky, Terkildsen, & Mayer, 2019). Although physical experiments play a key role in science learning (e.g. Haury & Rillero, 1994) and some research has already studied the combination of physical and virtual laboratories (see Jong, Linn, & Zacharia (2013) for a review of this topic), little is known about integrating multimedia learning in inquiry-based learning processes with physical experiments in real classroom settings (Oliveira et al., 2019).

In this regard, mobile digital devices such as smartphones and tablet PCs open up new possibilities for integrating multimedia learning, especially in science education. Due to their integrated sensors, they can measure physical data and visualize them via applications as representations such as tables, graphs etc. automatically and in real time. In this way, they can be used as a digital experimental tool to investigate scientific phenomena by augmenting experimental learning environments with multiple (external) representations (MERs). While Sung, Chang, and Liu's (2016) meta-analysis provided evidence for positive effects on learning performance for the instructional use of mobile digital media, they also claim that a blanket statement on how mobile digital media can be meaningfully used in the classroom is not possible due to the wide variety of possible teaching and learning scenarios as well as media and learning programs used. Rather, each newly developed digitally supported learning process must be tested for the fit of the medium and the learning program, as well as the learning effectiveness. In addition, Zydney and Warner (2016) concluded in their review report on mobile apps for science learning that future studies are needed to better align underlying theories, design, and outcome measures.

Following this conclusion, initial studies have shown positive effects of using mobile devices to augment experimental learning with MERs on conceptual learning (e.g. Becker, Klein, & Kuhn, 2018; Becker, Klein, Gößling, & Kuhn, 2019; Klein, Kuhn, & Müller, 2018) and motivation (e.g. Hochberg, Kuhn, & Müller, 2018). In order to further close the research gap, the present study aims to empirically investigate whether augmentation with MERs can effectively support the experimental learning process in real classroom settings for the use of a specific digital experimental tool, the tablet PC-supported video analysis.

## 1.1 Learning with MERs using video motion analysis in physics education

The important role of MERs for scientific learning is well documented for the natural sciences in general (Tytler, Prain, Hubber, & Waldrip, 2013), and for physics in particular (Treagust, Duit, & Fischer, 2017). It is of especially great importance for conceptual understanding (Verschaffel, de Corte Erik, Ton, & Jan, 2010) and discussed as a necessary condition for deeper understanding (diSessa, 2004). Ainsworth (2006, 2008) provides a broad overview of the unique benefits of MERs for learning complex or new scientific content. According to her, DeFT (Design, Functions, Tasks) taxonomy, learning with MERs means that two or more external representations are used simultaneously (e.g., diagrams, formulas and data tables). Helping students acquire representational competences is an important educational goal in many STEM (Science, Technology, Engineering, and Mathematics) domains. In particular, students need to acquire connection-making competences: they need to conceptually understand how different representations map to one another, and they need to be perceptually fluent in translating between representations (Rau, 2017). One promising method for fostering representational competence is video motion analysis, a method for non-contact measurement of time and position of moving objects frame by frame commonly used for computer-based experimental learning in physics and sports. From a video-based recording of moving objects, their velocity and acceleration can be calculated. The measurement data can then be visualized in diagrams and tables. Furthermore, markers in the video or graphs in sync next to the video image can be positioned (see **Figure 1**).

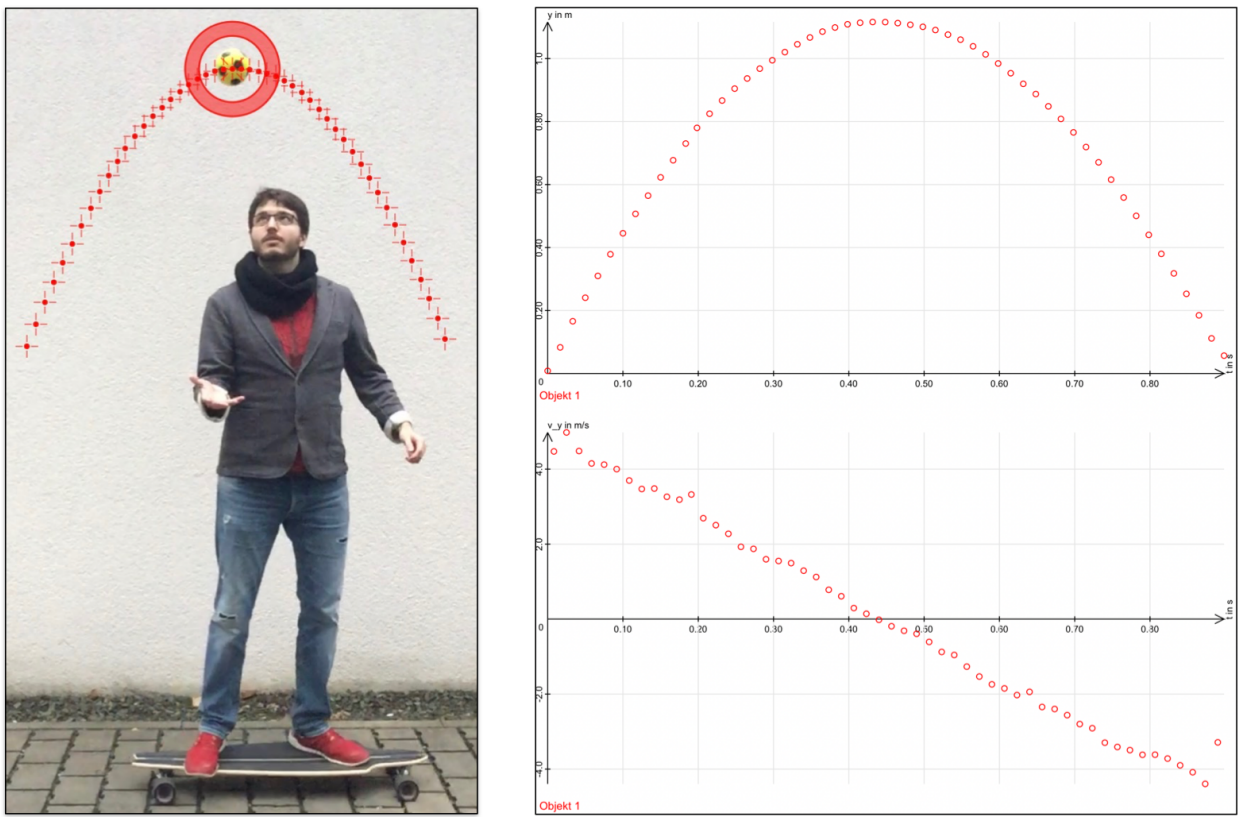

***Figure 1.*** Screenshots from video motion analysis application: video with tracking of object (left), y(t)-(position-time) graph (upper right), v_y(t)-(velocity-time) graph (lower right)

Tablet PCs nowadays have technically advanced cameras that can record videos also of (fast) moving objects in outstanding quality. Video analysis applications developed especially for physics lessons create the possibility of combining all of the single video analysis process steps on one and the same mobile device, from the recording of the moving object in an experiment to the analysis of the movement to the visualization of the relevant physical quantities by providing MERs such as tables, strobe pictures and diagrams. The learner can then switch between these MERs as needed and view them in combination.

In this way, video-based experimental tasks can promote in particular the interpretation and application of MERs and could be integrated in the DeFT framework (Ainsworth, 2006). Ainsworth proposes three key functions of MERs for supporting the learning process: According to the first function, MERs can complement each other either by providing complementary information or by allowing for complementary approaches to processing information. Video analysis fulfills this function by offering the learner different forms of representation for the analysis of one and the same movement (e.g., stroboscopic imaging and the corresponding motion diagrams). The second function is that simultaneously presented representations can constrain each other's interpretation in two ways: the more familiar representations can constrain the interpretation of the less familiar one, or inherent properties of one representation can trigger the usage of the other representation. Video analysis fulfills this function by displaying the real motion sequence, which the learners are familiar with, and simultaneously the stroboscopic imaging, a form of representation the learners are less familiar with. According to the third function, MERs can lead to a deeper understanding by allowing integration of corresponding information of the provided representations. In this way, video analysis supports the construction of deeper understanding by decomposition of a two dimensional into the associated one dimensional movements and simultaneously providing the learner with the time-position and time-velocity graphs.

Positive effects for the use of video analysis in physics education could be empirically demonstrated with regard to different forms of representation (Beichner (1998) for diagrams, Boyd & Rubin (1996) for strobe pictures, Pappas, Koleza, Rizos, & Skordoulis (2002) for tables and Kanim & Subero (2010) for vectors) as well as conceptual understanding in the field of mechanics, both for the introductory phase of physics as well as high school physics classes (Klein et al., 2018; Becker, Klein, & Kuhn, 2018; Becker et al., 2019; Hockicko et al., 2014; Wee et al., 2015).

Even though MERs have the potential to promote learning processes, they also create complex demands and can even increase cognitive load for the learners. Indeed, there are many studies pointing towards students difficulties with MERs (e.g. Ainsworth, 2006; Nieminen, Savinainen, & Viiri, 2010). Consequently, the cognitive load of the learning environment must be considered and managed carefully to enable successful learning with MERs.

## 1.2 Managing Cognitive Load in Multimedia Learning Environments

Basic assumption of the Cognitive Load Theory (CLT; Sweller, 1988; van Merriënboer & Sweller, 2005) is that working memory has a limited capacity not only in terms of the amount of information that can be processed simultaneously but also in terms of the time at which information is available for processing. In the relevant literature (e.g. Leppink & van den Heuvel, 2015; Sweller, van Merriënboer, & Paas, 2019), there are three types of cognitive load: intrinsic cognitive load (ICL), extraneous cognitive load (ECL) and germane cognitive load (GCL). ICL refers to the complexity of the information the learner must process during the learning process and is therefore determined by the learning task as well as the prior knowledge of the learner regarding the learning content. ECL refers to learning-irrelevant cognitive processes, which occupy the working memory but do not lead to relevant learning gain and can be influenced by the design of the learning procedure (e.g., how the information is presented to the learner). GCL refers to the amount of cognitive resources that are needed while processing the information in a learning process. Based on this fundamental theory, one fundamental learning principle is to keep ECL as low as possible during the learning process (Leppink, 2017; Leppink & van den Heuvel, 2015). In particular, when learning with (multiple) external representations, this can be achieved by avoiding the split-attention effect (Mayer & Pilegard, 2014; Sweller et al., 2019). This (negative) learning effect implies that the separation of related information sources requires mental integration processes, thus increasing ECL and inhibiting the learning process by occupying mental resources which are no longer available for learning-related processing.

The Cognitive Theory of Multimedia Learning (CTML; Mayer, 2005) builds on CLT's principle of a limited working memory capacity. According to Mayer, one feature of effective multimedia instruction should be to "reduce extraneous processing, manage essential processing, and fostering generative processing" (Mayer, 2009, p. 57). To help guide instructional designers, Mayer identifies 12 instructional principles. One of them is the principle of contiguity (Mayer & Moreno, 2003), which aims at reducing extraneous processing by avoiding the split-attention effect. It means that corresponding information should not be presented spatially or temporally separated from the learner (Mayer & Moreno, 2003; Mayer, 2009). The use of video analysis applications fulfills this principle by simultaneously presenting the learner different forms of corresponding representations in combination (e.g., stroboscopic imaging and the corresponding motion diagrams or time-position and corresponding time-velocity diagram). One further principle in the context of dynamic visualizations is the segmentation or interactivity principle (Mayer & Pilegard, 2014), which postulates a learning-conducive effect when learners themselves can determine the sequence or tempo of the information presentation. Video analysis applications allow the learners themselves to control the transition between the individual forms of representation. For example, to improve their understanding of the motion diagrams, the students can again call up the stroboscopic image. According to the segmentation principle, this self-control of the learning process avoids cognitive overload.

## 1.3 The present study

Physical conceptual understanding is considered a fundamental condition factor both for understanding physical learning contents (Vosniadou, 2007) as well as the ability to solve physical problems (Rittle-Johnson, Siegler, & Alibali, 2001). According to Pundak and Rozner (2007), there is a growing consensus that traditional teaching methods are insufficiently supportive of students' understanding of physical concepts. As reported in the introduction, video analysis applications can promote the learning process by fostering interpretation and application of MERs and reduce ECL by fulfilling the principle of contiguity and the segmentation principle.

Although the learning effectiveness of video motion analysis has been demonstrated in several studies, especially in the field of mechanics (e.g. Hockicko et al., 2014; Wee et al., 2015), the theoretical assumptions that account for the learning effects have yet not been investigated explicitly. The present study contributes to closing this research gap by empirically investigating and searching for connections between cognitive load and the learning effectiveness regarding physical conceptual understanding for a specific learning scenario in regular school lessons. For this purpose, we conducted a cluster-randomized controlled trial in a pre-post test design involving high school physics courses.

## 2 Research Hypothesis and Research Questions

From the theoretically-founded positive influence of video analysis applications on the learning process, we derive the following hypothesis:

*Research Hypothesis 1*

Compared to traditional teaching, augmented experimental learning environments based on tablet PC-supported video analysis lead to a reduction of ECL.

*Research Hypothesis 2*

Compared to traditional teaching, augmented experimental learning environments based on tablet PC-supported video analysis lead to a better conceptual understanding of the addressed physical concepts.

In CLT/CTML, it is assumed that with a reduction of ECL, more cognitive resources are available to the learner for active knowledge construction. This should lead to a more efficient learning process resulting in an increased learning performance. If differences in ECL and learning gain could be found in favor of technology-supported compared to traditional teaching sequences, it is unclear if the reduction in ECL is really causal for the enhanced learning gain, which leads to the following research question.

*Research Question*

Can a causal connection between the reduction of ECL and the learning gain be statistically supported?

# 3 Method

## 3.1 Sample

The data were collected in 18 courses from 11 secondary schools in different states in Germany between 2017 and 2018. In each experimental and control condition, questionnaires were completed in the students' regular classrooms in the presence of the associated teacher. In total, 294 students participated in both test times, that is above the desired sample size of 252, derived from a priori power-calculation using the software G*Power ($1$-$\beta$=0.999) based on the effect sizes found in our preliminary studies (Authors 2018; Authors 2019). Sociodemographic data were evaluated for 286 students, of whom 94 are female and 191 are male (one did not complete the question about gender), with an average age of 15.6 ($SD$=0.72). To ensure the comparability of the groups, only physics advanced courses were selected for the study. The sociodemographic composition of the population separated according to TG and CG is shown in **Table 1**.

**Table 1**

| Variable | TG ($N$=150) | CG ($N$=136) |
|---|---|---|
| Average age | 15.6 | 15.7 |
| Female (in %) | 34.0 | 31.9 |
| AC physics (in %) | 100.0 | 100.0 |
| AC mathematics (in %) | 38.9 | 37.7 |

*Sociodemographic composition*

*Note*: Data basis $N$=286.

**3.2 Study design**

The study should take place in a natural teaching-learning situation in which the students are taught together in the course organization. Since the spatial separation of the students of one course was not possible due to the general conditions at the participating schools, we followed Dreyhaupt et al. (2017) and opted for a cluster randomization and randomly assigned whole courses as treatment group (TG) or control group (CG). The study covered a curricular valid, essential topic of mechanics, the uniform motion. The teachers involved voluntarily participated in the study, and the students had already spent several weeks of physics teaching together in the respective courses before the beginning of the study. The lesson sequence focused on the fostering of physical conceptual understanding by gaining insights through independent, collaborative experimentation in small groups of two students. The effect of the intervention is captured by the operationalization of two dependent variables: cognitive load and physical conceptual understanding.

**3.3 Instruments**

**3.3.1 Teacher behavior.** The teacher's behavior during the intervention was evaluated on a 5-item scale for post-timing control of whether the learners were sufficiently supported by their teacher. For this purpose, a questionnaire was used, which has already been evaluated in our preliminary study (Authors 2018; Authors 2019). The items aimed to measure the commitment, willingness to support, and motivating effect of the teacher.

**3.3.2 Conceptual understanding.** In order to determine the learning gain regarding conceptual understanding (CU) for the experimental learning process and thus to empirically test the validity of the research hypothesis, a multiple-choice performance test was used (see

Supplementary Material), which consists of adapted items from validated test instruments (KCT (Lichtenberger, Wagner, Hofer, Stern, & Vaterlaus, 2017), KiRC (Klein, Müller, & Kuhn, 2017) and TUG-K (Beichner, 1994)) as well as self-developed items and has already been evaluated in our preliminary study (Authors, 2018; Authors, 2019). The test is structured in three in physics education approved sub-concepts: "Velocity as alteration rate" (G1), "Velocity as vectorial quantity" (G2) and "Reference system" (G3). The items (three per sub-concept) contain the common forms of representations used in kinematics: diagram, table, and strobe picture. To confirm the intended structuring, we performed a factor analysis on the CU response pattern at the post-time point (see Appendix).

**3.3.3 Cognitive load.** A ten-item subjective survey developed and validated by Leppink, Paas, van der Vleuten, van Gog, and van Merriënboer (2013) was used to measure the intervention-induced cognitive load. With this instrument, it is possible to not only measure the overall cognitive load, but also differentiate between the three types of cognitive load (Hadie & Yusoff, 2016; Zukić, Đapo, & Husremović, 2016): ICL, ECL and GCL. We follow Leppink, Paas, van Gog, van der Vleuten, and van Merriënboer (2014) and interpret GCL as a subjective judgment of learning and, as suggested by Leppink and van den Heuvel (2015), use a two-factor cognitive load model which incorporates only intrinsic and extraneous cognitive load for subsequent analyses. Since the original questionnaire was developed for students in a statistics course, the items had to be adjusted to the specific physical context in this study (see Supplementary Material) and literally translated into German. Even though empirical studies have already demonstrated that the three-factor model is robust against adaption to the disciplinary context (Leppink et al., 2014), we decided to verify the three-factor structure for the adapted questionnaire used in this study by factor analysis (see Appendix).

### 3.4 Experimental Manipulation

Desktop-based video analysis has the disadvantage that recording and analysis of moving objects within experiments are temporally separated. Since this interrupts the learning process and complicates the implementation in regular school lessons, as time on task is extended compared to the conducting of the experiment using conventional methods, we decided to use tablet PCs and the video analysis application Viana[1]. While students in TG conducted and recorded experiments with a tablet PC, analyzed motion processes of these experiments, and visualized the measuring data with different MERs, students in the CG conducted, recorded, analyzed and visualized experimental motion processes with different MERs using experimental tools established in traditional school education: a stopwatch, tape measure and graphing calculator. In order to allow a fair comparison between the TG and the CG, the experiments, learning content, time on task, as well as social form of learning were identical. In particular, the students in both groups used the same forms of representation for learning: diagram, strobe picture, table and formula (**Table 2** compares the number of individual forms of representation used for learning during the intervention for both groups). The students in both groups were given learning tasks for conducting the experiments and analyzing the measured data developed for this study in cooperation with teachers with many years of professional experience. To illustrate the comparability, **Figure 2** shows an example learning task that the students of the TG and the CG had to work on. In the 4x45-minute lesson sequence, the students experimented independently in pairs. The experimental learning process of each learning sequence (2x45 minutes) in each group involved the set-up, execution, and evaluation of two experiments. For the (low-cost) experimental setup, the students used only an aluminum profile and a steel sphere. The experiments aimed at the fostering of physical conceptual understanding regarding the uniform motion and were designed so that the students could carry out the experiments completely independently, thus reducing the influence of the teacher as far as possible. For this reason, the learning gain results from the independent processing of the learning tasks and not from interaction with the teacher. Moreover, the involved teachers were instructed to take a passive role during the experimentation process rather than actively intervene. However, they were allowed to respond to inquiries regarding experiment set-up and execution. This was to ensure that all student groups were able to carry out the experiments successfully.

---

1 The application Viana is available for iOS for free at https://goo.gl/4RWv8g, a detailed description of the application is available from Becker et al. (2018).

**Table 2**

| Form of representation | TG | CG |
| --- | --- | --- |
| Diagram | 9 | 8 |
| Strobe picture | 2 | 2 |
| Table | 1 | 4 |
| Formula | 5 | 5 |

*Number of forms of representation used during the intervention*

*Note*: Data basis *N*=286.

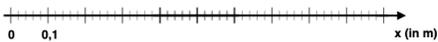

**Figure 2.** Exemplary learning task for the students of the CG (left) and the students of the TG (right)

## 3.5 Study procedure

In the first lesson of the intervention, the students of the TG received standardized instruction in the physical video analysis with the tablet PC. The guidance included an explanation of the measurement methodology and the video analysis application used. In addition, the students were given the opportunity to familiarize themselves with the video analysis application by analyzing sample videos. The sample videos are already included in the library of the application and have no contextual relationship to the subject area of the study. The students of the CG received standardized instruction in data analysis with the graphing calculator and were given the opportunity to get familiar with the functionality of the graphing calculator by analyzing given data sets, also with no contextual relationship to the subject area of the study.

In the first lesson after the introductory lesson, the sociodemographic data as well as the preliminary marks in physics, mathematics and German were requested. Next, the students completed the pre-performance test on conceptual understanding under the supervision of the associated teacher. During the next four lessons, the students conducted the experiments. In the immediately following lesson, a post-test was carried out under the supervision of the associated teacher. In order to maximize the comparability, the post-test included identical items, but in a different order. Following the post-test, students were asked to complete the cognitive load questionnaire as well.

## 4 Results

### 4.1 Preliminary analysis

**4.1.1 Covariate balance.** In order to examine if the population is balanced at the pre-time point regarding preliminary marks and foreknowledge, a Wilcoxon-Mann-Whitney test (U test) was carried out. The results show a significant difference between the groups for the preliminary mark in physics and mathematics as well as the conceptual knowledge regarding uniform motion (see **Table 3**).

**Table 3**

| Variable | TG *M (SD)* | CG *M (SD)* | U-test *p* |
|---|---|---|---|
| Preliminary mark | | | |
| Physics | 2.26 (1.20) | 2.63 (1.41) | 0.040 |
| Mathematics | 2.54 (1.25) | 3.01 (1.37) | 0.004 |
| German | 3.05 (1.09) | 3.25 (1.15) | 0.100 |

| Conceptual understanding | | | |
|---|---|---|---|
| Pre-test score | 4.66 (2.09) | 3.92 (1.99) | 0.004 |

*Test for significant group differences*

*Notes*: Data basis *N*=286. For marks, 1=very good, 2=good, 3=satisfactory, 4=sufficient, 5=poor, 6=deficient.

Since it can not be excluded that these covariates can affect the outcome of the intervention, a sample balanced in these covariates was generated with the method of Propensity Score Matching (PSM; Rosenbaum and Rubin,1983; Guo & W. Fraser, 2010) prior to the following comparative analysis. PSM allows causal statements to be made about intervention effects in empirical studies in which complete randomization is not possible or sufficiently successful from the outset (Fan & L. Nowell, 2011). Based on a logistic regression model of all potentially confounded variables, the propensity score (PS) for each subject is estimated from the total population as probability of belonging to one of the comparative groups. Each participant in one group is assigned one or more participants of the other group with the same or very similar values of the PS. Following that, the treatment effect can be estimated for the matched population with conventional statistical techniques.

To match the given population, all covariates collected (preliminary marks, pre-test score) were used to determine the PS. We decided on the matching technique known as "nearest neighbor matching", which matches a student of the TG to a student of the CG that is closest in terms of a distance measure estimated by logistic regression. After the matching process, the population is balanced in all covariates (see **Table 4**), but the sample size has been reduced from $N$=286 to $N$=262. However, this sample size is still sufficient for further statistical analysis, so that all further evaluations are based on the data set of the balanced sample.

**Table 4**

|  | TG | CG | U-test |
|---|---|---|---|
| Variable | *M (SD)* | *M (SD)* | *p* |
| Preliminary mark |  |  |  |
| Physics | 2.34 (1.20) | 2.61 (1.40) | 0.157 |
| Mathematics | 2.71 (1.22) | 3.00 (1.37) | 0.101 |
| German | 3.05 (1.13) | 3.24 (1.16) | 0.157 |
| Conceptual understanding |  |  |  |
| Pre-test score | 4.31 (1.94) | 3.95 (2.01) | 0.198 |

*Test for significant group differences after PSM*

*Notes*: Data basis $N$=262. For marks, 1=very good, 2=good, 3=satisfactory, 4=sufficient, 5=poor, 6=deficient.

**4.1.2 Teacher behavior.** It was checked whether teacher behavior had an influence on the learning performance of the students. To this end, a correlation analysis was performed which provided, in accordance with the results of our preliminary study (Authors, 2018; Authors, 2019), no indication of a correlation between teacher behavior and learning gain, neither for the entire population nor the sub-groups. As a result, the teacher behavior is not taken into account in the subsequent analyses.

### 4.2 Analysis of cognitive load data

**Figure 3** gives an overview of the group-dependent averages and standard errors of the different sub-scales. In order to find significant group differences among the different types of intervention-induced cognitive load, the student scores of the sub-scales ICL, ECL, and GCL were subjected to a one-factorial analysis of variance (ANOVA). Before the analysis, it was checked if the assumptions for conducting ANOVA had been met (independence of samples, normal distribution of residuals and homogeneity of residuals' variances). The results are presented in **Table 5**. In this way, a significantly lower intervention-induced extraneous cognitive load for the TG could be detected with high test power ($F(1,240)=27.01$, $p<10^{-3}$, $\eta^2=0.101$, $1\text{-}\beta=1.000$).

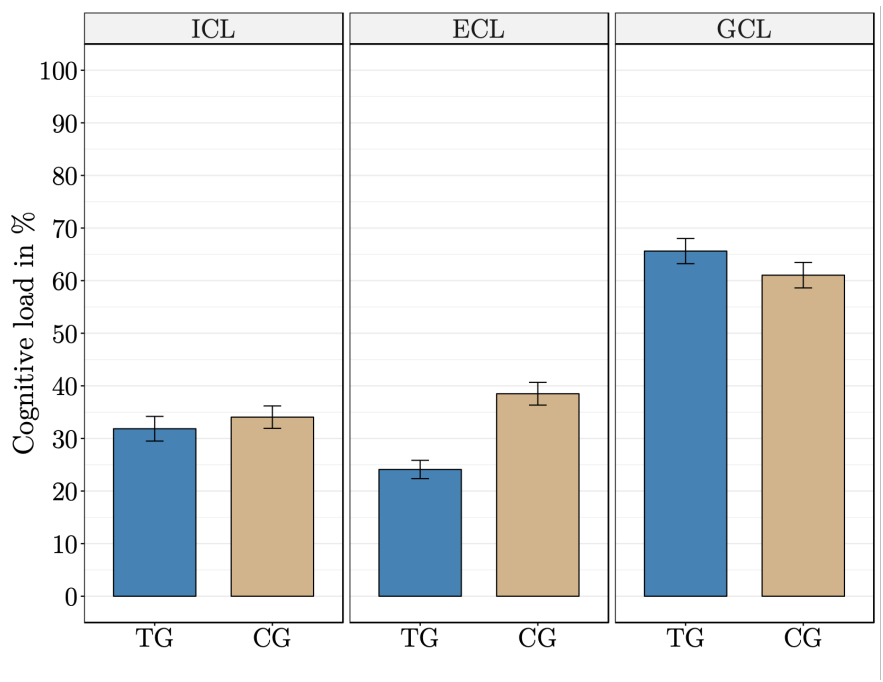

***Figure 3***. Intervention-induced cognitive load

**Table 5**

| Sub-scales | $F(1,240)$ | $p$ | $\eta^2$ | $1\text{-}\beta$ |
|---|---|---|---|---|
| ICL | 0.101 | 0.751 | - | - |
| ECL | 27.01 | $<10^{-3}$ | 0.101 | 1.000 |
| GCL | 0.156 | 0.694 | - | - |

*Results of ANOVA*

*Note*: Data basis *N*=262.

### 4.3 Analysis of performance data

For each item, students were asked to rate their confidence on a 4-point Likert scale ranging from "very sure" to "guessed." The confidence information will be used at this point to rate guessed answers post-hoc as incorrect, which contributes to the validity and reliability of the test results. **Figure 4** gives an overview of the group-dependent averages and standard errors of relative test scores separated according to the different sub-concepts. To reveal significant group differences among the learning gain, the performance data was subjected to a repeated measures ANOVA (rmANOVA) with treatment as the between subject effect and CU test score as the repeated measure. Before the analysis, it had been checked if the assumptions for conducting rmANOVA are met (independence of samples, normal distribution of residuals, and sphericity). The results are presented in **Table 6**. As a result of the analysis, a significant group difference could be found for sub-concept G3 in favor of the TG with high test power ($F(1,259)$=10.82, $p$=0.001, $\eta^2$=0.048, $1\text{-}\beta$=0.953).

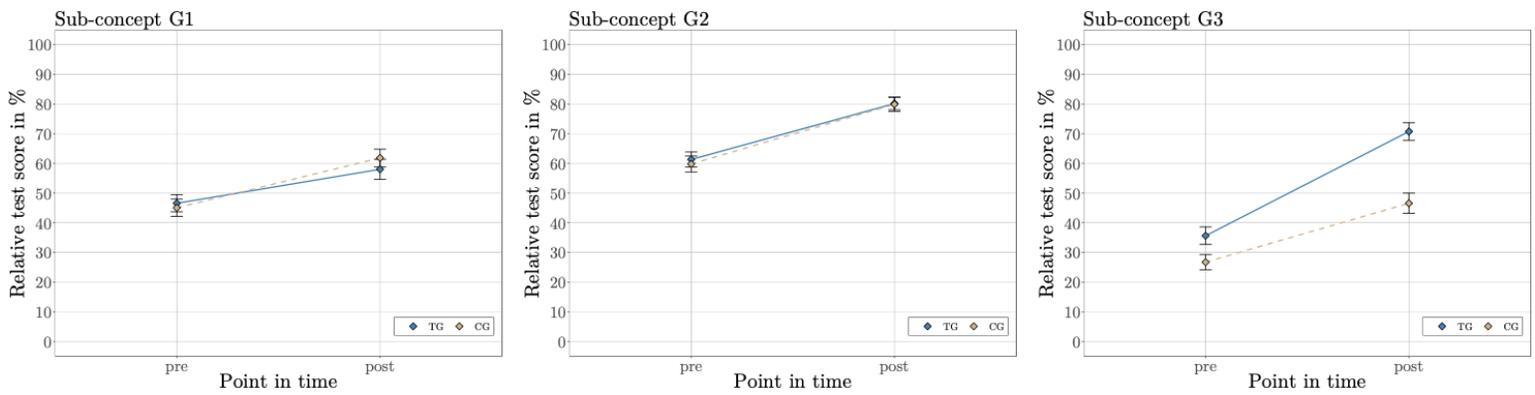

*Figure 4*. Relative test scores for sub-concepts

**Table 6**

| Sub-concept | $F(1,260)$ | $p$ | $\eta^2$ | $1-\beta$ |
|---|---|---|---|---|
| G1 | 1.656 | 0.199 | - | - |
| G2 | 0.105 | 0.746 | - | - |
| G3 | 10.82 | 0.001 | 0.048 | 0.953 |

*Results of rmANOVA*

*Note*: Data basis *N*=262.

### 4.4 Structural equation modeling

To investigate whether the theoretically-founded causal relation between the reduction of ECL and the increase in learning gain can be statistically supported, we used structural equation modeling (SEM; Simon, 1951) using the R-Package lavaan (vs. 0.6-3). It is agreed upon in the scientific community that a large sample size is required for sufficient statistical power of SEM. Following the methodological recommendations in the authoritative literature (e.g. Kline, 2011; Barrett, 2007), the sample size in our study is above the lower limit of 200, so we consider the sample size to be appropriate. The measurement model derived and approved by factor analysis (see Appendix) for the cognitive load variables (ICL & ECL) and the cognitive performance variables (sub-concepts G1, G2 & G3) was fit into a structural model to clarify the influence of cognitive load on learning performance. Since the data is ordinally scaled and not multivariate normally distributed, the diagonally-weighted least squares procedure was used to estimate the model parameters. The underlying structural model is presented in **Figure 5**. The resulting model with 58 free parameters fit the data acceptably well: $p(\chi^2)$=0.048, *CFI*=0.937, *TLI*=0.964, *RMSEA*=0.035, *SRMR*=0.040. The significant path coefficients of the regression model are reported in **Table 7**

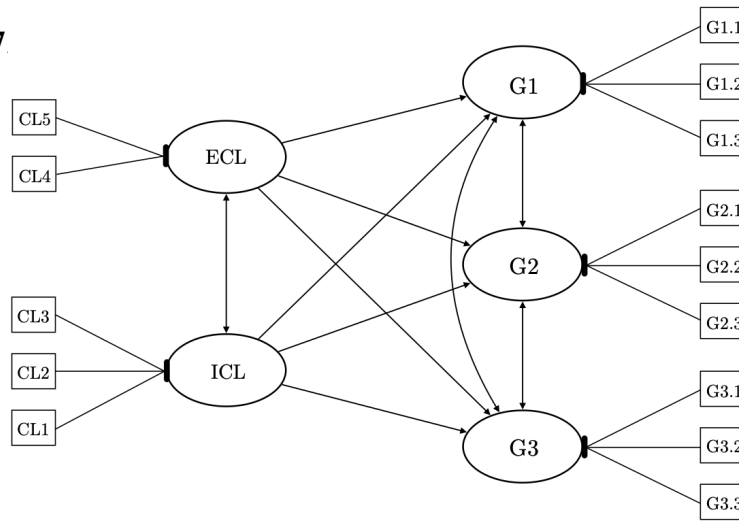

***Figure 5***. Structural equation model

**Table 7**

| Exogenous variables | Endogenous variables | *b* | *se* | *z* | *p* | *β* |
|---|---|---|---|---|---|---|
| ICL | G1 | -0.317 | 0.138 | -2.292 | 0.022 | -0.248 |
| ICL | G2 | -0.304 | 0.152 | -2.001 | 0.045 | -0.348 |
| ECL | G3 | -0.514 | 0.170 | -3.029 | 0.002 | -0.463 |

*Significant path coefficients*

*Notes*: Data basis *N*=241, *b* = unstandardized estimate, *β* = standardized estimate.

## 5 Discussion

The cluster-randomized controlled trial presented in this work was aimed at the empirical investigation of the effectiveness of augmenting the experimental learning process with MERs using tablet PC-supported video analysis in a real classroom setting. The digital experimental tool was used in regular high school lessons for an essential topic in mechanics instruction, uniform motion. It could be shown that the augmentation with MERs using the tool leads, in comparison to a control group taught traditionally, to a significant reduction of ECL and also had a positive impact on conceptual understanding of the description of movement with regard to a reference system. Moreover, a statistical analysis of causality related to the connection between ECL reduction and increased conceptual understanding empirically supported the theoretical assumptions of the method's learning effectiveness.

### 5.1 Effects on cognitive load

While ICL and thus the complexity of the learning content is comparable for both groups, the intervention-induced ECL is significantly lower for the students of the TG. This supports our research hypothesis that the automatic visualization of MERs in combination (which fulfills the contiguity principle and thus avoids the split attention effect) and the possibility for the learners to switch between them as needed without time delay (which fulfills the segmentation principle) contributes to a reduction of ECL. This insight also provides an explanation for previous positive research findings (Klein et al., 2018; Becker, Klein, & Kuhn, 2018; Becker et al., 2019) and proves that augmenting the experimental learning process with MERs using a mobile device can successfully support the students during the experimentation process. In summary, it can be stated that the positive effects of the video analysis on cognitive load remain with the implementation in regular school lessons using a mobile device.

## 5.2 Effects on conceptual understanding

A positive effect of the augmentation of the experimental learning process with MERs using video motion analysis on the development of physical conceptual understanding was found for sub-concept G3. This confirms the results of our own preliminary study (Author, 2018; Author, 2019) and the research findings from Hockicko et al. (2014) and Wee et al. (2015), who have already been able to empirically demonstrate a positive effect of the video analysis on conceptual understanding in other topics in mechanics. Since sub-concept G3 refers to analysis of the movement in dependence on the reference system, one possible cause for the positive effect found is that the video analysis application allows the learner to interact actively with the representation of the coordinate system, to manipulate it, and to analyze its effects on other representations in real time. This means that the application enables the learner to determine the origin and spatial orientation of the coordinate system independently and to change it as needed (which fulfills the segmentation principle). The effects of the change of the coordinate system on the movement diagrams can then be observed quasi-simultaneously, which means that the learner can display the associated diagram by a gesture of the hand, without noticeable time delay. On the one hand, this reduces the split-attention effect, and on the other hand, it increases the learners' active interaction with MERs and thus active knowledge construction. In summary, the results support our research hypothesis, but only for one out of three sub-concepts.

**5.3 Connection between cognitive load and conceptual understanding**

Using the method of SEM, the causal relationship between cognitive load variables and cognitive performance variables has been statistically demonstrated. The regression modeling delivers a significant negative path coefficient for ECL with sub-concept G3, for which the significant group difference in favor of the TG occurs, while for ICL with the other sub-concepts. Following from this, it can be assumed that learners only need assistance through augmentation with MERs in understanding sub-concept G3 and that a reduction of ECL is actually the cause of an enhanced learning gain of the TG. That supports our research hypothesis that the learning effectiveness of the video analysis is based on the reduction of cognitive load. This also shows that the consideration of the fundamental learning theories (CLT, CTML, DeFT) and the recommendations derived from them are of great importance for the success of regular teaching scenarios in everyday school life.

**5.4 Practical Implications**

First, we were able to implement the video motion analysis using a mobile device as a digital experimental tool in regular physics lessons with little training required for the participating teachers and students. This should encourage teachers to use this tool in their own lessons. Moreover, the additional effort is limited, since only one lesson has to be invested for the introduction of the tool. Second, in our study, we have shown that augmentation with MERs is particularly conducive to understanding uniform motion relative to a reference system in an experimental learning process. This should be considered by teachers when planning experimental learning environments on this topic. Since uniform motion is an essential, but relatively simple topic, the results of our study also suggest that the positive effects of video analysis on more complex topics in mechanics could have an even greater impact on the effectiveness of the learning process. This opens up new possibilities for teachers to create innovative experimental learning environments in physics lessons.

**5.5 Limitations**

To be able to generalize the results of our study to other contexts, we used a large sample size of students and conducted a cluster-randomized controlled trial by randomly assigning complete courses to TG or CG. To enable a fair comparison, the experiments and the learning time were identical for both groups. Moreover, the learning material in terms of learning content, representations used, and level of difficulty was comparable. However, in interpreting the study's findings, some limitations must be considered. First of all, it should be noted that due to the quasi-experimental design, a complete randomization could not be achieved, and the sample was unbalanced regarding the preliminary marks and the pre-test score. A balance regarding these variables was achieved using the PSM method, but this reduced the sample size and thus the statistical significance of the study. Second, the learning-related effect, based on the assessment used, was demonstrated for a special lesson scenario covering a specific topic of mechanics. Therefore, it remains unclear if the positive effects of the tablet PC-supported video analysis also hold for a more complex topic. Given this research deficit and the practical significance of the use of digital experimental tools, we hope that research groups will build on our results and conduct further studies to broaden the research findings. Third, the intervention covered a learning time of only four school lessons. Thus, the results of the study do not allow any conclusion on the learning effectiveness of the video analysis for a long-term teaching use. Again, we hope that our findings lead to further research, since video analysis may support the experimental learning process in numerous sub-topics of mechanics in regular school education. Fourth, although the sample size is large for an implementation study ($N$=262 after PSM), it is not much larger than an adequate sample size of $N$=200 (Kline, 2011) for SEM, indicating that the results of SEM should be replicated in a follow-up study with a larger sample size.

**5.6 Conclusion**

We demonstrated firstly that the augmentation of an experimental learning process with MERs using a mobile device in a real classroom setting reduces the learners' ECL significantly. Second, we revealed by comparing the learning performance with traditionally-taught students that the augmentation with MERs also leads to a deeper understanding of an important sub-concept of a curricular-valid, essential topic of mechanics, uniform motion. Third, we statistically verified that the reduction of ECL is the cause of this enhanced learning gain. The research results presented in this work thus contribute to explaining the positive effects found in preliminary studies for the use of the video analysis method in teaching-learning scenarios on the basis of fundamental theories of learning (CLT, CTML, DeFT). We continue to believe that reducing cognitive load leads to even greater effects on more complex topics of mechanics, demonstrating the potential of the method to sustainably improve experimental learning processes in regular school lessons. With regard to the increasing digitization in the education sector, we see this as productive research desiderata for future research in this area.

Appendix

***Factor analysis of CU performance test:*** Subsequent to an examination of the necessary conditions (KMO, Bartlett's test of sphericity), we performed an exploratory factor analysis (EFA) on the CU response pattern at the post-time point by using the R-package psych (vs. 1.8.12). In this way, we found an underlying factor structure according to the three sub-concepts (Kaiser criterion, Scree-Plot, Parallel Analysis) and tested the quality of this model by a confirmatory factor analysis (CFA) by using the R-package lavaan (vs. 0.6-3). The resulting model fits the data well, $p(\chi^2)$=0.251, *CFI*=0.994, *TLI*=0.991, *RMSEA*=0.022, *SRMR*=0.038.

***Factor analysis of CL questionnair:*** Subsequent to an examination of the necessary conditions (KMO, Bartlett's test of sphericity), we performed an EFA with a subsequent CFA. As expected, we confirmed the three-factor structure (Kaiser criterion, Scree-Plot, Parallel Analysis) as intended by Leppink et al. (2013) for the questionnaire used. Only one item (item CL6) could not be assigned to the factor structure, so we decided to remove this item from the subsequent analysis. Later, we tested the quality of the model with the remaining nine items by CFA. The resulting model fits the data well, $p(\chi^2)$=0.111, *CFI*=0.993, *TLI*=0.990, *RMSEA*=0.037, *SRMR*=0.023.

CL Questionnaire

CL1: The topic "accelerated motion" was very complex.

CL2: The formulas for "accelerated motion" I perceived as very complex.

CL3: The physical concepts and definitions of the topic "accelerated motion" I perceived as very complex.

CL4: The tasks and work assignments were very unclear.

CL5: The tasks and work assignments were, in terms of learning, very ineffective.

CL6: The tasks and work assignments were full of unclear terms.

CL7: Editing the tasks has really enhanced my understanding of the topic "accelerated motion".

CL8: Editing the tasks has really enhanced my understanding of the physics of "accelerated motion".

CL9: Editing the tasks has really enhanced my understanding of the formulas for "accelerated motion".

CL10: Editing the tasks has really enhanced my understanding of physical concepts and definitions of "accelerated motion".

**Information:** A body moves along a straight line at constant velocity from left to right.
**Question:** Which coordinate system must be chosen to give the following time-position diagram?

☐ Coordinate system 1
☐ Coordinate system 2
☐ Coordinate system 3
☐ Coordinate system 4

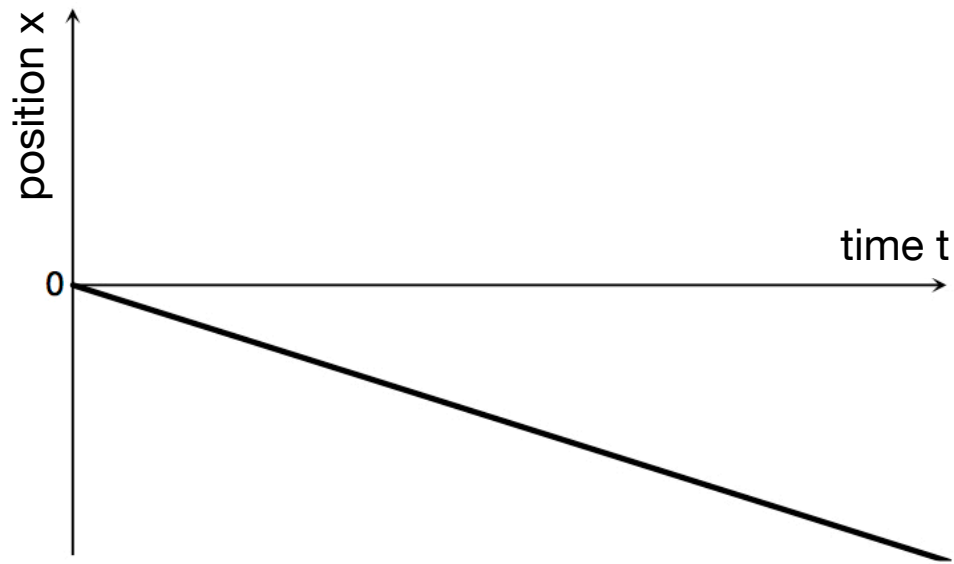

coordinate system 1

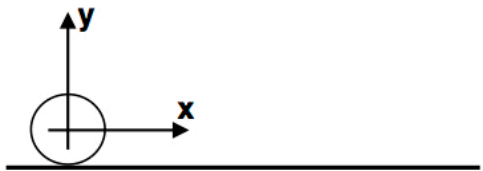

coordinate system 2

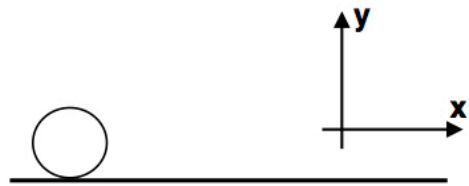

coordinate system 3

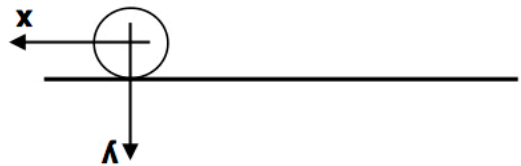

coordinate system 4

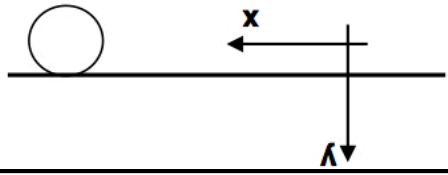

**Information:** A body moves along a straight line at constant velocity from left to right.

**Question:** Which coordinate system must be chosen to give the following time-position diagram?

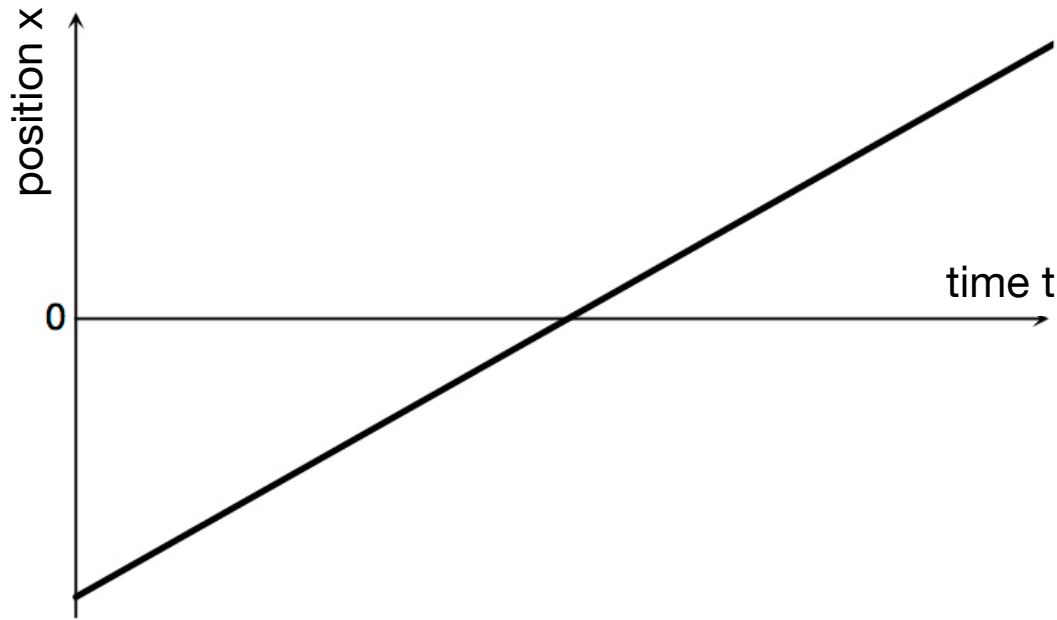

- ☐ Coordinate System 1
- ☐ Coordinate System 2
- ☐ Coordinate System 3
- ☐ Coordinate System 4

coordinate system 1

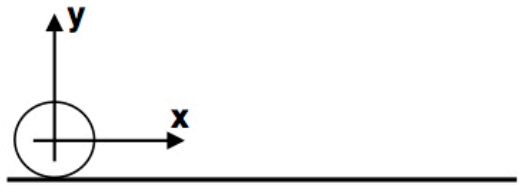

coordinate system 2

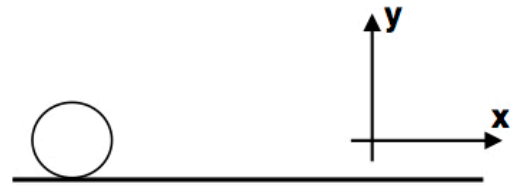

coordinate system 3

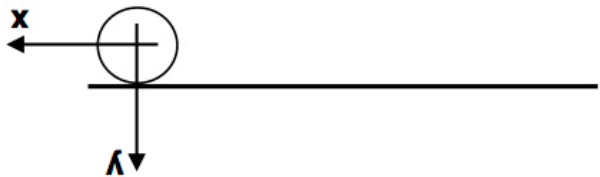

coordinate system 4

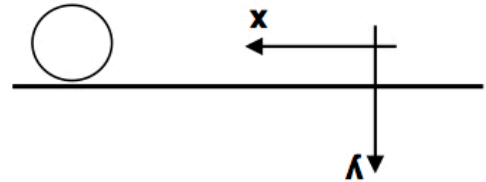

**Information:** Two bodies move along the same line from left to right. The figure shows the positions of the two bodies at equal time intervals at times 1 to 7.

**Question:** Do the bodies have the same velocity at some point?

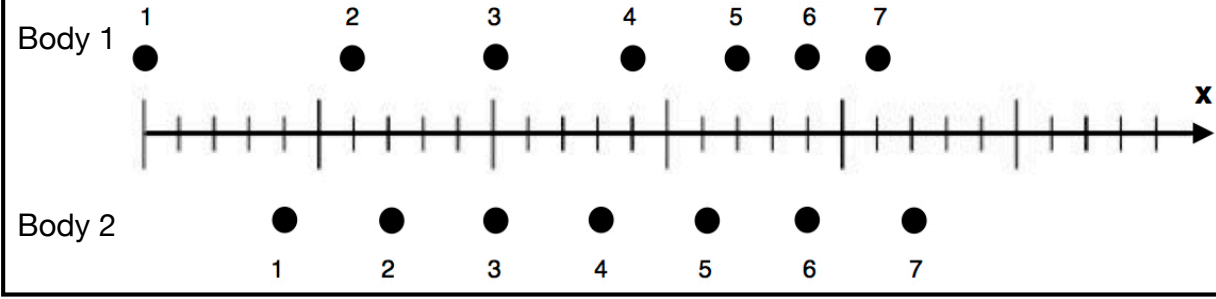

- ☐ Yes, sometime between the times 4 and 5
- ☐ Yes, at time 3
- ☐ Yes, at times 3 and 6
- ☐ No

**Information:** The diagram represents the movement of a body.
**Question:** What statement can be made about the movement of the body?

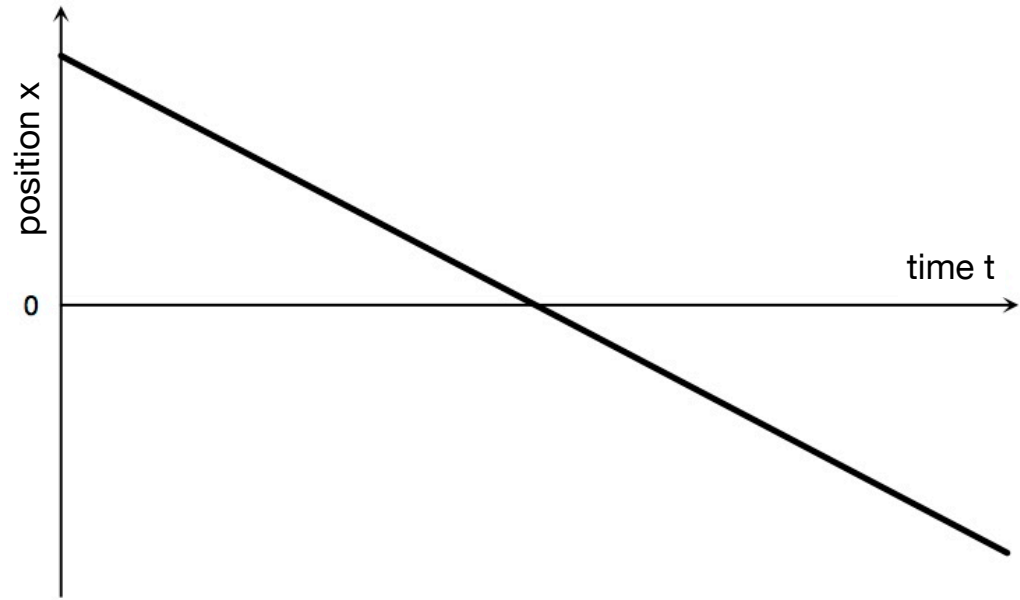

- [ ] The body is always moving forward
- [ ] The body always moves backwards
- [ ] The body first moves forward and then backwards
- [ ] The body is moving down an incline

**Information:** A body moves forward for 4s at a constant velocity and then moves backwards for 4s at half the velocity.

**Question:** Which table fits the movement of the body?

☐ Table 1
☐ Table 2
☐ Table 3
☐ Table 4

### Table 1

| time (in s)        | 0 | 2 | 4  | 6    | 8  |
|--------------------|---|---|----|------|----|
| position x (in m)  | 0 | 5 | 10 | 12,5 | 15 |

### Table 2

| time (in s)        | 0 | 2 | 4  | 6     | 8   |
|--------------------|---|---|----|-------|-----|
| position x (in m)  | 0 | 5 | 10 | -12,5 | -15 |

### Table 3

| time (in s)        | 0 | 2 | 4  | 6   | 8 |
|--------------------|---|---|----|-----|---|
| position x (in m)  | 0 | 5 | 10 | 7,5 | 5 |

### Table 4

| time (in s)        | 0 | 2  | 4   | 6     | 8   |
|--------------------|---|----|-----|-------|-----|
| position x (in m)  | 0 | -5 | -10 | -12,5 | -15 |

**Information:** 3 bodies move along the same line. The following diagram shows the movement of the 3 bodies.

**Question:** Which body has the highest velocity at time $t_1$?

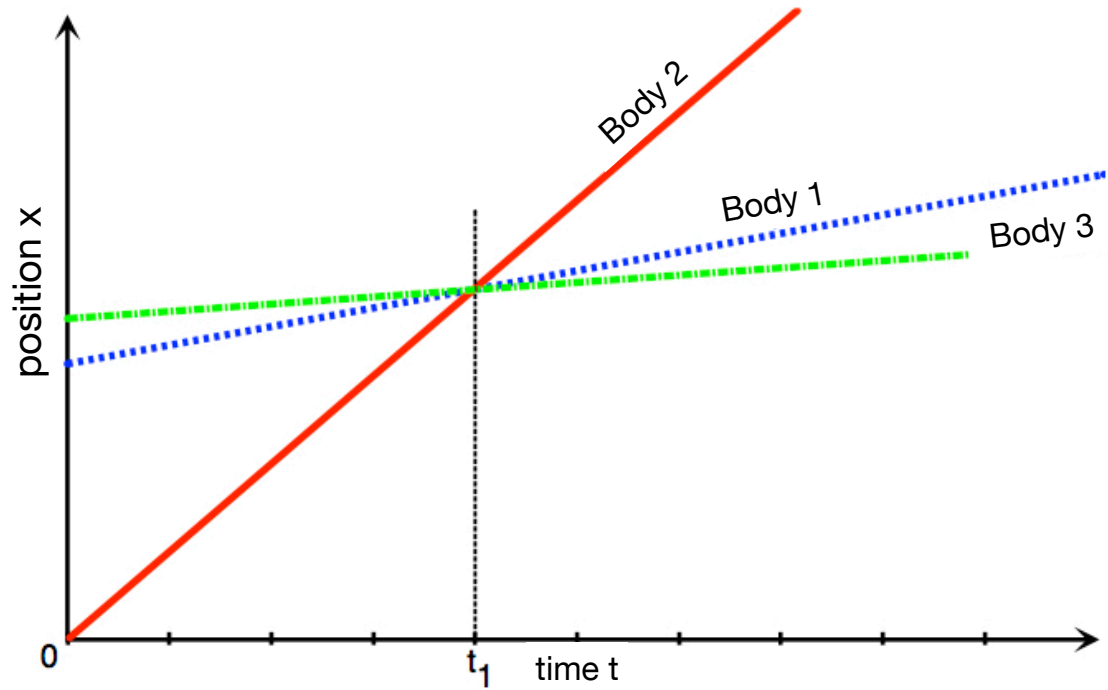

☐ Body 3

☐ Body 2

☐ All three bodies have at time $t_1$ the same velocity

**Information:** A body moves at a constant velocity along a straight line from left to right. The positions of the body at different times are listed in the following table.

**Question:** How should the coordinate system be chosen so that the given table results?

| Zeit (in s) | 0 | 1 | 2 | 3 | 4 |
|---|---|---|---|---|---|
| Ort x (in m) | 0 | -5 | -10 | -15 | -20 |

☐ Coordinate system 1
☐ Coordinate system 2
☐ Coordinate system 3
☐ Coordinate system 4

coordinate system 1

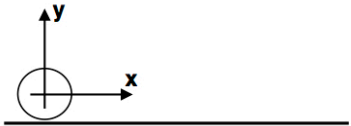

coordinate system 2

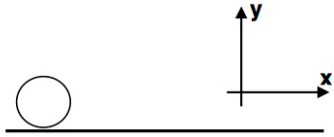

coordinate system 3

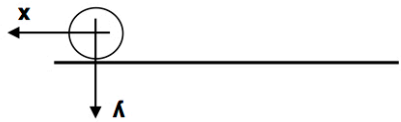

coordinate system 4

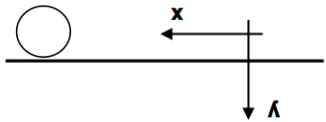

**Information:** Body 1 and Body 2 each move at constant velocity as shown in the opposite directions.
**Question:** What statement can be made about the signs of the velocities?

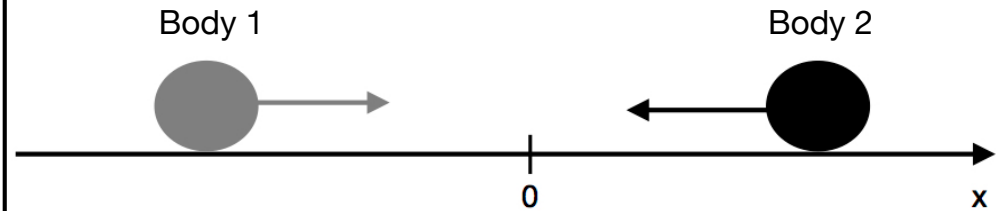

☐ The velocities of both bodies are positive

☐ The velocity of body 2 is positive, the velocity of body 1 is negative

☐ The velocity of body 1 is positive, the velocity of body 2 is negative

☐ About the sign of the velocities you can make no statement

**Information:** Two bodies move along the same line. The positions of the two bodies at different times are listed in the following table.

**Question:** Do the bodies have the same velocity at some point?

| time (in s) | 0 | 2 | 4 | 6 | 8 | 10 | 12 | 14 |
|---|---|---|---|---|---|---|---|---|
| Position x of body 1 (in m) | 2 | 4 | 7 | 11 | 16 | 22 | 29 | 37 |
| Position x of body 2 (in m) | 0 | 4 | 8 | 12 | 16 | 20 | 24 | 28 |

- ☐ No
- ☐ Yes, at the times 2s and 8s
- ☐ Yes, at the time 8s
- ☐ Yes, sometime between the times 4s and 6s